\documentclass[twocolumn, pra]{revtex4}
\usepackage{amsmath}
\usepackage{amssymb}
\usepackage{graphicx}
\newcommand{\mbf}[1]{\mathbf{#1}}
\newcommand{\ot}{\mathrm{otherwise}}
\newcommand{\vect}[1]{\mathbf{#1}}

\newtheorem{theorem}{Theorem}

\newtheorem{definition}{Definition}
\newtheorem{corollary}{Corollary}

\begin{document}

\title{ Strong nonlocality: A trade-off between states and measurements }

\author{Anthony J. Short$^1$}\email{ajs256@cam.ac.uk}
\author{Jonathan Barrett$^2$}

\affiliation{$^1$ DAMTP, Centre for Mathematical Sciences,
Wilberforce Road, Cambridge CB3 0WA, UK} \affiliation{$^2$
H.~H.~Wills Physics Laboratory, University of Bristol, Tyndall
Avenue, Bristol BS8 1TL, UK }

\begin{abstract}
Measurements on entangled quantum states can produce outcomes that
are nonlocally correlated. But according to Tsirelson's theorem,
there is a quantitative limit on quantum nonlocality. It is
interesting to explore what would happen if Tsirelson's bound were
violated. To this end, we consider a model that allows arbitrary
nonlocal correlations, colloquially referred to as ``box world''. We
show that while box world allows more highly entangled states than
quantum theory, measurements in box world are rather limited. As a
consequence there is no entanglement swapping, teleportation or
dense coding.
\end{abstract}

\maketitle

\section{Introduction}

Despite its great explanatory and predictive power, the standard
formalism of quantum theory - in which states are represented by
vectors in a complex Hilbert space - retains an abstract
mathematical character. Of course there is no reason why Nature
should not be described by an abstract mathematical formalism. But
it is notable that in quantum theory textbooks, the formalism is
simply postulated. It is not, say, derived from a small set of
elementary physical considerations in the manner of special
relativity. This invites the question: why \emph{that} structure, as
opposed to any other?

One way to approach this question, or at least gain some insight, is
to compare and contrast quantum theory with other models -
theoretical possibilities which do not describe our universe, but
which can nonetheless be explored. In this paper, we investigate one particular non-classical, non-quantum theory. In \cite{barrett}, this theory was called \emph{generalized non-signalling theory}, or GNST for short, as it admits all non-signalling correlations \cite{popescu, boxes}. Here we call it \emph{box world}.

One of the notable features of box world is that, as in quantum
theory, measurements on separate but entangled systems can produce
outcomes that are nonlocally correlated, i.e., which cannot be
explained by any local hidden-variable model \cite{bell, CHSH}. It
is already known that nonlocal correlations are useful in many
information theoretic tasks \cite{qi1, qi2}. But in quantum theory,
there is a quantitative limit on the amount of nonlocality that
correlations can have \cite{tsirelson}. In box world, by contrast,
arbitrary nonlocal correlations can be produced, as long as they do
not permit instantaneous signalling. This has consequences. For
example, with stronger than quantum correlations, it is known that
communication complexity problems become trivial, requiring only
constant communication \cite{van-dam,brassardetal}. On the other
hand, as we show in this paper, possibilities for measurement in box
world are in some ways rather limited. It turns out that there is
nothing analogous to a Bell measurement. We also show that there is
no entanglement swapping, teleportation, or dense coding. This
extends to the whole of box world previous special-case results in
\cite{barrett, coupler}.

As entanglement swapping is possible in the quantum world
\cite{swapping}, it follows that box world does not contain quantum
theory as a special case, and that it cannot be the true theory
describing reality.

\section{A framework for probabilistic theories}

In order to compare classical theories, quantum theories, and
alternatives such as box world, we need a common mathematical
framework in which the different theories can be written down. We
begin by describing such a framework. It is operational in flavour.
This means, for example, that \emph{system}, \emph{preparation} and
\emph{measurement} are all taken as basic terms. Different ways of
preparing a system will prepare different \emph{states}. The state
of a system determines the outcome probabilities for any measurement
that can be performed on the system. The framework we describe and
the notation we use is closely based on that of \cite{lucien} and
\cite{barrett}. However, we should note that there is nothing very
novel in the framework itself, and that numerous formalisms have
been developed over the years, intended to be operational
generalizations of classical and quantum theory (see e.g.,
\cite{lucien, dariano, convex1, convex2}).

\subsection{Single systems}

One way of specifying the state of a system would be to give an
exhaustive list of the outcome probabilities for every possible
measurement. However, as Hardy points out \cite{lucien}, most
physical theories have enough structure that it is not necessary to
give such an exhaustive list in order to specify the state fully.
Instead, for each type of system, assume that its state can be
completely characterised by the outcome probabilities for some
finite subset of all possible measurements. Following Hardy, we
refer to the subset chosen to represent the state as \emph{fiducial}
measurements. If the outcome probabilities for the fiducial
measurements are known, then the state is known, and the outcome
probabilities for any other measurement can be inferred. For
example, the state of a single qubit in quantum theory corresponds
to a density operator on a 2-dimensional Hilbert space. But equally,
it can be completely characterised by the outcome probabilities of
measurements corresponding to the three Pauli operators, $\sigma_x,
\sigma_y$ and $\sigma_z$. Note that the choice of which measurements
to use as fiducial measurements is not unique.

A convenient way of writing down the state is as a vector
$\vect{P}$, with components $P(a|x)$, where this is the probability
of obtaining outcome $a$ when fiducial measurement $x$ is performed.
Obviously, $P(a|x)$ must be positive and normalised such that
$\sum_{a} P(a|x) = 1$. For example, for binary valued $a$ and $x$:
\begin{equation}
\vect{P} = \left(\begin{array}{c} P(0|0) \\ P(1|0) \\ \hline P(0|1)
\\ P(1|1) \end{array} \right).
\end{equation}

Within a particular operational model, it is not necessarily the
case that any vector $\vect{P}$ represents an allowed state (that
is, a state which can actually be prepared). In quantum theory there
is no qubit state that assigns probability 1 to the +1 outcome for
all three Pauli measurements. An operational model must specify, for
each type of system, a set $\mathcal{P}$ of states which can
physically be prepared. We assume that arbitrary probabilistic
mixtures of states can be prepared (e.g., by tossing some coins and
preparing either state $\vect{P}$ or state $\vect{Q}$ depending on
the result). Hence $\mathcal{P}$ is convex.

\subsection{Multi-partite systems}

Most of the interesting questions that can be asked in information
theory involve more than one system. So how should we describe the
state of a multi-partite system in a general operational model?
Consider $n$ systems, $A_1,\ldots,A_n$, with a set of fiducial
measurements for each. If a fiducial measurement $x_1$ is performed
on $A_1$, $x_2$ on $A_2$, and so on, then at the least, the joint
state of $A_1,\ldots,A_n$ should determine a joint probability for
each combination of outcomes. We assume something further: a
specification of the joint probability of each combination of
outcomes for each possible combination of fiducial measurements is
sufficient to determine completely the joint state. Note that this
property does indeed hold in both quantum theory and classical
probability theory. But it is not trivial. For example, in an
alternative quantum theory, formulated using real instead of complex
Hilbert spaces, the property does not hold \cite{realqm1, realqm2,
realqm3}.

It is convenient to write a multi-partite state of $n$ systems in
the form of an $n$-dimensional array, with entries
$P(\mbf{a}|\mbf{x})$, where $\mbf{x}=(x_1, \ldots, x_n)$ specifies a
fiducial measurement for each subsystem, $\mbf{a}$ is a list of the
corresponding measurement outcomes, and $P(\mbf{a}|\mbf{x})$ is the
probability of getting $\mbf{a}$ given $\mbf{x}$. For example, for
two systems, each with a binary measurement choice $x$, and binary
outcomes $a$:
\begin{equation}
\vect{P} = \left( \begin{array}{cc|cc} P(00|00) & P(01|00) & P(00|01) & P(01|01) \\
P(10|00) & P(11|00) & P(10|01) & P(11|01) \\ \hline
P(00|10) & P(01|10) & P(00|11) & P(01|11) \\
P(10|10) & P(11|10) & P(10|11) & P(11|11) \end{array} \right)
\end{equation}
Conditions of positivity and normalisation apply as before.

Finally, all multi-partite states must satisfy the
\emph{no-signalling conditions}:
\begin{quote}
$\sum_{a_n} P(\mbf{a}|\mbf{x})$ \; is independent of $x_n$ for all
$n$.
\end{quote}
These ensure that separated parties cannot send messages to one
another simply by making measurements on their subsystems. Arguably,
if the no-signalling conditions do not hold, then we had no right to
be speaking of separate subsystems in the first place.

A multi-partite state is a \emph{product state} if $\vect{P}$
satisfies
\begin{equation}
P(\mbf{a}|\mbf{x}) = P_1(a_1|x_1)P_2(a_2|x_2)\ldots P_n(a_n|x_n),
\end{equation}
where $P_i(a_i|x_i)$ is a valid state of the $i$th system. A state
is \emph{separable} if it can be written as a convex combination of
product states, otherwise it is \emph{entangled}.

\subsection{Measurements} \label{sec:measurement}

In general, the fiducial measurements will not be the only
measurements that one can perform on a system. For example, on a
single qubit, there is a measurement corresponding to
$\sigma_{45^{\circ}} = 1/\sqrt{2} (\sigma_x + \sigma_z)$. On two
qubits there is a Bell measurement. By the definition of the
fiducial measurements, it must be possible to derive the measurement
probabilities for all such measurements from the
$P(\mbf{a}|\mbf{x})$.

In fact, by considering mixtures of states, it can be shown that the
probability $\mathrm{Pr}(r)$ of obtaining a particular outcome $r$
in any measurement must be a linear function of the fiducial
measurement probabilities \cite{lucien,barrett}. We can therefore
associate an \emph{effect} $\vect{R}_r$ with each measurement
outcome, where $\vect{R}_r$ is an array with components
$R_r(\mbf{a}|\mbf{x})$, such that
\begin{equation}
\mathrm{Pr}(r) = \vect{R_r} \cdot \vect{P} \equiv \sum_{\mbf{a x}}
R_r(\mbf{a}|\mbf{x}) P(\mbf{a}|\mbf{x}).
\end{equation}
A measurement with various possible outcomes is associated with a
set $\{ \vect{R}_r  \}$. For example, consider again a qubit in
quantum theory, with fiducial measurements chosen to be $\sigma_x,
\sigma_y, \sigma_z$. The measurement corresponding to
$\sigma_{45^{\circ}}$ is represented by the $1$-dimensional arrays
\begin{equation} \label{eqn:example_meas}
\vect{R}_{+1} =\left(\begin{array}{r} 2^{-\frac{3}{2}} \\ - 2^{-\frac{3}{2}} \\ \hline 2^{-1} \\
2^{-1} \\ \hline 2^{-\frac{3}{2}} \\ - 2^{-\frac{3}{2}}
\end{array} \right),\ \ \ \vect{R}_{-1} =\left(\begin{array}{r} -2^{-\frac{3}{2}} \\ 2^{-\frac{3}{2}} \\ \hline 2^{-1} \\
2^{-1} \\ \hline -2^{-\frac{3}{2}} \\  2^{-\frac{3}{2}}
\end{array} \right)
\end{equation}
Note that the array $\vect{R}_r$ associated with a measurement
outcome is not in general unique, since there may be a different
array $\vect{R}'_r$ satisfying $\vect{R}_r\cdot\vect{P} =
\vect{R}'_r\cdot\vect{P} \; \forall \, \vect{P} \in \mathcal{P}$.

A particular operational model must contain a specification of the
set of measurements that can physically be performed on a particular
type of system. There is a constraint: any measurement must
correspond to a set $\{\vect{R}_r\}$ such that
\begin{equation}\label{constraint1}
\vect{R}_r\cdot\vect{P} \geq 0 \ \ \forall r\ \forall
\,\vect{P}\in\mathcal{P}
\end{equation}
and
\begin{equation}\label{constraint2}
\sum_r \vect{R}_r\cdot\vect{P} = 1 \ \ \forall
\,\vect{P}\in\mathcal{P}.
\end{equation}
Furthermore, if a measurement is performed on one subsystem of a
bipartite system, it is possible to calculate the subsequent
(``collapsed'') state of the other subsystem, conditioned on a
particular outcome. Clearly this should be an allowed state of that
subsystem.

\subsection{Dynamics}\label{dynamics}

In addition to preparations and measurements, it may be possible to
perform transformations on a system, i.e., to act on it in such a
way that the system is preserved but its state changes. In the most
general case, a system can change into a system of a different type.
But here we consider only transformations that preserve the type of
system. Such a transformation can be represented as a map
$\vect{T}:\mathcal{P}\rightarrow\mathcal{P}$. As with measurements,
a consideration of mixed states implies that $\vect{T}$ is linear.
Thus a transformation can be represented by an array such that
\begin{equation}\label{dynamicsequation}
P'(\mbf{a}'|\mbf{x}') = \sum_{\mbf{a x}} T(\mbf{a}'|\mbf{x}',
\mbf{a}|\mbf{x}) P(\mbf{a}|\mbf{x}).
\end{equation}
For each type of system, an operational model should specify a set
of physically possible transformations. A valid transformation
should satisfy
\begin{equation}
\vect{T}(\vect{P}) \in \mathcal{P} \ \forall \vect{P} \in
\mathcal{P}.
\end{equation}
There are other consistency conditions that the sets of allowed
states, measurements and transformations should satisfy, which we
will not go into in detail. For example, if a transformation is
followed by a fiducial measurement, then this process taken as a
whole should correspond to a valid measurement on the initial state.

\section{Box World}

Box world is one particular operational model, which has a natural
definition in terms of the framework defined above, and which it is
interesting to compare with the quantum and classical theories. Box
world is defined as follows. Any $\vect{P}$ satisfying:
\begin{enumerate}
\item Positivity: $P(\mbf{a}|\mbf{x})\geq 0$
\item Normalisation: $|\vect{P}| = \sum_{\mbf{a}} P(\mbf{a}|\mbf{0}) =1 $
\item No-signalling: $\sum_{a_n} P(\mbf{a}|\mbf{x})$ is independent of $x_n$
\end{enumerate}
is an allowed state.\footnote{In \cite{barrett}, for simplicity, the
number of possible outcomes for each fiducial measurement are taken
to be the same. However, all the results presented in this paper
also apply (without modification)  to the more general case in which
each fiducial measurement may have a different number of outcomes.}

Two subtleties: first, the normalisation condition is only stated
for the measurement choice $\mbf{x}=\mbf{0}$. But the no-signalling
conditions are then sufficient to ensure normalisation for all
measurement choices since they imply
\begin{equation}
\sum_{\mbf{a}} P(\mbf{a}|\mbf{x})=\sum_{\mbf{a}}
P(\mbf{a}|\mbf{x'}).
\end{equation}
Second, although the no-signalling conditions refer only to fiducial
measurements, it can be shown that they are sufficient to prevent
signalling using any kind of measurement \cite{barrett}.

Box world permits many states that do not have counterparts in
quantum theory. An interesting example is this bipartite state:
\begin{equation} \label{PS-state_eqn}
P_{PR} (a_1 a_2 | x_1 x_2) = \left\{
\begin{array}{ccl} \frac{1}{2} & : & a_1
+ a_2 = x_1 x_2\; \textrm{(mod 2)} \\
0& :  & \ot
\end{array} \right. ,
\end{equation}
where $x_1, x_2, a_1, a_2 \in\{0,1\}$. The correlations generated by
performing fiducial measurements on this state are nonlocal, meaning
that they violate the Clauser-Horne-Shimony-Holt (CHSH) inequality
\cite{CHSH}. In fact, they are \emph{more nonlocal} than is possible
in quantum theory, because they return a value of 4 for the CHSH
expression. Tsirelson's theorem \cite{tsirelson} shows that quantum
correlations always return a value $\leq 2\sqrt{2}$, and CHSH showed
that local correlations always return a value $\leq 2$. These
superquantum nonlocal correlations have appeared in the literature
before \cite{popescu, boxes}, and they are sometimes referred to as
a Popescu-Rohrlich (PR) box. Thus we refer to this state as the PR
box state.

What are the allowed measurements in box world? The model is defined
so that any set $\{\vect{R}_r\}$ satisfying conditions
(\ref{constraint1}) and (\ref{constraint2}) above corresponds to a
physically possible measurement. In the following, we explore what
kinds of measurement this actually allows, and the consequences for
information processing. Intuitively, the fact that conditions
(\ref{constraint1}) and (\ref{constraint2}) must be satisfied means
that there is a tradeoff between states and measurements. If there
is a larger space of states, then there is a smaller set of
measurements that are compatible with those states. The space of
states in box world is in an obvious sense maximal, so one would
expect the possibilities for measurement to be less interesting than
in, say, quantum theory. This is indeed the case.

\section{Measurements in box world}

The following property of measurements in box world is proven in
Appendix D of \cite{barrett}.
\begin{theorem}\label{positivity}
All effects in box world can be represented using only positive
arrays where each component satisfies
\begin{equation}
0 \leq R(\mbf{a} | \mbf{x}) \leq 1.
\end{equation}
\end{theorem}
From hereon, assume that effects are indeed represented this way.

In the case of multi-partite systems, we have already defined
product states, and distinguished separable and entangled states.
Similar definitions apply to effects. A multi-partite effect is a
\emph{product effect} if
\[
R(\mbf{a}|\mbf{x}) = R_1(a_1|x_1)R_2(a_2|x_2)\ldots R_n(a_n|x_n),
\]
where $R_i(a_i|x_i)$ is a valid effect on the $i$th system. An
effect is separable if it can be written as a sum of product
effects, otherwise it is entangled. (Note that the sum here really
is just a sum - not a convex combination, as in the definitions
applicable to states.) Any array $\vect{R}$ with one entry $\in
(0,1]$ and the rest zero represents a product effect. Hence
Theorem~\ref{positivity} yields
\begin{corollary}\label{noenteffect}
There are no entangled effects in box world.
\end{corollary}
Section~\ref{consequences} shows that Theorem~\ref{positivity} also
prevents entanglement swapping and teleportation in box world.

\begin{figure}
\includegraphics{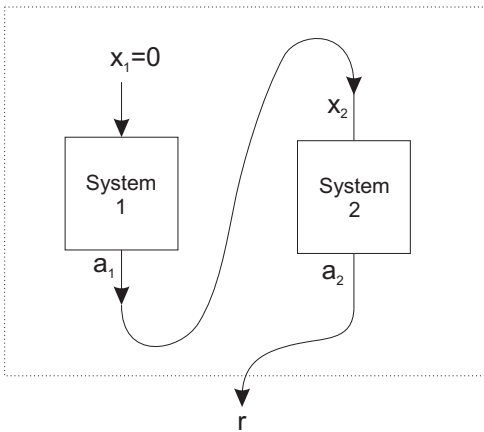}
\caption{An example of a basic measurement on two systems, each with
$x,a \in \{0,1\}$. } \label{fig:basic_measurement}
\end{figure}

A certain class of measurements is particularly simple and will play
a special role in what follows.
\begin{definition}
A measurement is \emph{basic} if it can be implemented by a sequence
of fiducial measurements on individual subsystems, where later
measurement choices may depend (deterministically) on earlier
outcomes, and the final measurement outcome $r$ is given by a
deterministic function of the fiducial measurement outcomes
$\mbf{a}$.
\end{definition}
An example of a basic measurement for two subsystems is given in
Figure~\ref{fig:basic_measurement}, where fiducial measurement
$x_1=0$ is performed on the first subsystem and then measurement
$x_2=a_1$ is performed on the second, and the final measurement
outcome is given by $r=a_2$.
In general, it is reasonable to require that measurements of this
form are in the set of allowed measurements in an operational model.
In fact, as we can also choose such basic measurements at random, it
is also reasonable to require that convex combinations of basic
measurements (defined in an obvious way) are allowed. Note that for
a single system, each basic measurement corresponds to a fiducial
measurement, possibly with relabelled outputs.

\begin{theorem} All valid measurements on single or bipartite systems in box world are convex combinations of basic measurements. \label{wiringtheorem}
\end{theorem}
This means that any measurement on a single or bipartite system can
be implemented by a probabilistic protocol involving only fiducial
measurements. Especially given Corollary~\ref{noenteffect}, it would
be natural to assume that Theorem~\ref{wiringtheorem} generalises to
multi-partite systems in box world, and indeed this was hypothesized
in Ref.~\cite{barrett}. However, things are not so simple.
\begin{theorem}
For tri-partite systems, there are measurements which do not reduce
to a convex combination of basic measurements.\label{3systemtheorem}
\end{theorem}
Section~\ref{counterexample} illustrates this with a specific
example.

Finally, there is at least some limitation on the power of
measurements in box world, even in the multi-partite case.
\begin{theorem}
For any single or multi-partite system, all allowed measurements can
be simulated using fiducial measurements and post-selection (i.e.,
the measurement is allowed to sometimes fail).
\label{postselecttheorem}
\end{theorem}

\section{Proofs}

Some preliminary remarks will be useful. First, recall
Theorem~\ref{positivity}, which states that the entries of an effect
$\vect{R}$ can be assumed non-negative. This is used throughout this
section.

Now consider the arrays $\vect{R}_r$ corresponding to the outcomes
of a basic measurement. Suppose that a basic measurement is carried
out, with final outcome $r$, and that during its execution, the
fiducial measurements $\mbf{x}$ are performed, with outcomes
$\mbf{a}$. In this case, say that the triple $\{r,\mbf{a},\mbf{x}\}$
is realized. A basic measurement can be represented such that the
component $R_r(\mbf{a}|\mbf{x})$ is equal to 1 if and only if it is
possible for $\{r,\mbf{a},\mbf{x}\}$ to be realized, else it is 0.
From hereon we make this choice. For example, the bipartite
measurement illustrated in Fig.~\ref{fig:basic_measurement} is
represented by
\begin{eqnarray}
R_0(a_1 a_2| x_1 x_2) &=& \left( \begin{array}{c|c}  \begin{array}{cc} 1&0 \\ 0&0 \end{array}  &   \begin{array}{cc} 0&0 \\ 1&0  \end{array} \\
\hline  \begin{array}{cc} 0&0 \\ 0&0 \end{array}  &
\begin{array}{cc} 0&0 \\ 0&0 \end{array} \end{array} \right),
\label{r0example}\\
R_1(a_1 a_2| x_1 x_2) &=& \left( \begin{array}{c|c}
\begin{array}{cc} 0&1 \\ 0&0 \end{array}  &
\begin{array}{cc} 0&0 \\ 0&1  \end{array} \\ \hline
\begin{array}{cc} 0&0 \\ 0&0 \end{array}  &
\begin{array}{cc} 0&0 \\ 0&0 \end{array}
\end{array} \right).\label{r1example}
\end{eqnarray}

\begin{definition}
The \emph{total measurement array} for a measurement  $\{ \vect{R}_r
\}$ is given by
\begin{equation}
\vect{M}= \sum_r \vect{R}_r.
\end{equation}
\end{definition}
From Eq.~(\ref{constraint2}), $\vect{M}$ satisfies
\begin{equation}\label{dynamic_eqn}
\vect{M} \cdot \vect{P} = 1\ \ \forall \vect{P} \in \mathcal{P}.
\end{equation}

Now consider the total measurement array corresponding to a basic
measurement. It has a simple form, which can be described
iteratively. First, since each $\mbf{a}$ corresponds to a specific
$r$, the component $M(\mbf{a}|\mbf{x})$ is equal to 1 if and only if
it is possible for the pair $\{ \mbf{a}, \mbf{x} \}$ to be realized.
The probability of $\{ \mbf{a}, \mbf{x} \}$ being realized is given
by
\begin{equation} \label{eqn:basic_meas_probs}
\mathrm{Pr}(\mbf{a}, \mbf{x}) = M(\mbf{a}|\mbf{x})\times
P(\mbf{a}|\mbf{x}).
\end{equation}
Now, for a single system, $M(a|x)=\delta_{xi}$ for some $i$. For a
multi-partite system composed of $n$ subsystems, there must exist
some $k$, and some $i$, such that the first step in the basic
measurement is to perform the fiducial measurement $x_k=i$ on the
$k$th subsystem. Let $\mbf{\hat{x}}_{ k }$ represent a sequence of
measurements and $\mbf{\hat{a}}_{ k }$ a sequence of outcomes on the
remaining $n-1$ subsystems. $\vect{M}$ satisfies
\begin{equation}
M(\mbf{a}|(x_k\neq i)  \mbf{\hat{x}}_{ k })=0.
\end{equation}
Define a new $(n-1)$-dimensional array $\vect{M}_{a_k}$ such that
\begin{equation}
M_{a_k} (\mbf{\hat{a}}_{ k }|\mbf{\hat{x}}_{ k }) \equiv M(a_k
\mbf{\hat{a}}_{ k }|(x_k=i)  \mbf{\hat{x}}_{ k } ).
\end{equation}
For all $a_k$, $\vect{M}_{a_k}$ must correspond to a valid basic
measurement on $(n-1)$ subsystems.

Finally, with a suitable choice of deterministic function of
$\mbf{a}$ for the output $r$, $M(\mbf{a}|\mbf{x})$ (consisting of 0s
and 1s) can be decomposed into any sum of arrays
$R_r(\mbf{a}|\mbf{x})$ (also consisting of 0s and 1s). For the
measurement of Figure~\ref{fig:basic_measurement}, outcomes are
represented by $R_0$ and $R_1$ as in Eqs.~(\ref{r0example}) and
(\ref{r1example}), and $\vect{M}$ is given by
\begin{equation}
M(a_1 a_2| x_1 x_2) = \left( \begin{array}{c|c}
\begin{array}{cc} 1&1 \\ 0&0 \end{array}  &
\begin{array}{cc} 0&0 \\ 1&1  \end{array} \\ \hline
\begin{array}{cc} 0&0 \\ 0&0 \end{array}  &
\begin{array}{cc} 0&0 \\ 0&0 \end{array}  \end{array} \right).
\end{equation}

\subsection{Proof of Theorem~\ref{wiringtheorem}}

Consider a measurement on $n$ systems with outcomes $\{\vect{R}_r\}$
and total measurement array $\vect{M}$. The measurement is a convex
combination of basic measurements if it can be performed by rolling
dice, say, and then performing one basic measurement or another
depending on the outcome of the dice roll. In this case it is
obvious that $\vect{M}$ is a convex combination of total measurement
arrays for basic measurements.

The first step in the proof of Theorem~\ref{wiringtheorem} is to
note that the converse also holds. That is, given $\vect{M}$ and
$\{\vect{R}_r\}$, if $\vect{M}$ can be written as a convex
combination of total measurement arrays for basic measurements, then
there is a convex combination of basic measurements with the same
total measurement array $\vect{M}$, and with outcomes corresponding
to $\{\vect{R}_r\}$. To see this, suppose that $\vect{M}=\sum_i q_i
\vect{M}_i$, where $0\leq q_i\leq 1$, $\sum_i q_i = 1$ and
$\vect{M}_i$ is the total measurement array for a basic measurement.
Construct an appropriate convex combination of basic measurements as
follows. With probability $q_i$, let the order of fiducial
measurements to perform be that dictated by $\vect{M}_i$. The
probability of measuring $\mbf{x}$ and obtaining outputs $\mbf{a}$
is given by Equation~(\ref{eqn:basic_meas_probs}) (which continues
to hold for convex combinations of basic measurements). In the case
that $\{ \mbf{a},\mbf{x} \}$ is realized, announce result $r$ with
probability $\mathrm{Pr}(r|\mbf{a}, \mbf{x} ) =
R_r(\mbf{a}|\mbf{x})/M(\mbf{a}|\mbf{x})$ (where if $\{
\mbf{a},\mbf{x} \}$ is realized, the right hand side must be $\geq
0$ and $\leq 1$). The overall probability of obtaining outcome $r$
is now given by
\begin{eqnarray}
\mathrm{Pr}(r) &=& \sum_{\mbf{a} \mbf{x}} \mathrm{Pr}(r|\mbf{a}, \mbf{x} ) \mathrm{Pr}(\mbf{a}, \mbf{x}) \\
&=&\sum_{\mbf{a} \mbf{x}} R_r(\mbf{a}|\mbf{x}) P(\mbf{a} | \mbf{x})
\\ &=& \vect{R}_r . \vect{P},
\end{eqnarray}
as required.

In order to prove Theorem~\ref{wiringtheorem}, it is thus sufficient
to show that for any measurement on a single or bi-partite system in
box world, $\vect{M}$ can be written as a convex combination of
total measurement arrays for basic measurements. Let a
\emph{subnormalised} basic measurement array have the form $\vect{M}
= \alpha \overline{\vect{M}}$, for $0 \leq \alpha \leq 1$ and
$\overline{\vect{M}}$ a total measurement array. The strategy is to
show that given a (non-zero) subnormalised $\vect{M}$, it is always
possible to subtract a (non-zero) subnormalised basic measurement
array $\vect{M}_B$ to leave a subnormalised
$\vect{M}'=\vect{M}-\vect{M}_B$ with at least one additional zero
entry. By iteration, we can then prove that any normalised
$\overline{\vect{M}}$ can be built from a convex combination of
basic measurement vectors.

Since the total measurement array $\vect{M}$ is central to the
analysis of this section, from here on we use the term `measurement'
to refer both to the measurement itself and to $\vect{M}$, depending
on context.

We will use variables with subscripts to denote sets of numbers,
indexed by the value of the subscript. For example, $a_x$ refers to
a set of $a$ values (one for each value of $x$). We will use
notation such as $x^*$ to refer to a particular $x$-value.

\subsubsection{Single-system measurements}

In Ref.~\cite{barrett} it is shown that all single-system
measurements in box world are mixtures of fiducial measurements. We
include an alternative proof here to illustrate the techniques used
in the bi-partite case in a simpler setting.

Consider a (non-zero) subnormalised $\vect{M}$ associated with a
single system characterised by any set of fiducial measurements. One
of the following must hold:

\begin{enumerate}

\item It is possible to subtract a (non-zero) subnormalised basic measurement $\vect{M}_B$ from $\vect{M}$ to leave a valid subnormalised  $\vect{M}' = \vect{M}-\vect{M}_B$. In this case there must be a measurement choice $x^*$ for which $M(a|x^*) >0 \; \forall \, a$, hence it is possible to subtract the subnormalised basic measurement $M_B(a|x) = k \delta_{x,x^*}$ where $k=\min_a (M(a|x^*))$. This generates at least one additional 0 entry in $\vect{M}'.$
\label{yeswiring}

\item It is not possible to subtract a (non-zero) subnormalised basic measurement from $\vect{M}$ without creating a negative component. In this case, there must exist a set $a_x$ such that $M(a_x |x)=0 \; \forall \, x$.   \label{nowiring} In terms of the representation of $\vect{M}$ as an array, there must be at least one zero in each block.

\end{enumerate}

Case~\ref{nowiring}, however, is impossible. Consider the state
$\vect{P}$ defined by $P(a|x) = \delta_{a, a_x}$. $\vect{P}$ is
clearly an allowed state, as it is positive, normalised and
non-signalling. However, if case~2 holds then
\begin{equation}
\vect{M}.\vect{P} = \sum_{ax} M(a|x) P(a|x) = 0.
\end{equation}
This implies that $\vect{M}.\vect{P} = 0$ for all states $\vect{P}$.
But the only way to achieve this is to take $\vect{M}= 0$, which
contradicts the initial assumption that $\vect{M}$ is non-zero.
Therefore case~\ref{yeswiring} is the only possibility.
Subnormalised basic measurements can be subtracted from $\vect{M}$
until the zero vector remains. Hence any single-system measurement
can be expressed as a finite sum of subnormalised basic
measurements. Since $\overline{\vect{M}}.\vect{P} = 1$ for any state
$\vect{P}$ and normalised $\overline{\vect{M}}$, this procedure
yields a decomposition of $\overline{\vect{M}}$ as a convex
combination of basic measurements.

\subsubsection{Bi-partite system measurements}

Consider a (non-zero) subnormalised $\vect{M}$, associated with a
bipartite system, with components $M(ab|xy)$. (In this subsection,
fiducial measurements are labelled $x, y$, and outcomes $a, b$,
rather than $x_1, x_2$ and $a_1,a_2$ respectively.) One of the
following must hold:

\begin{enumerate}

\item It is possible to subtract a (non-zero) subnormalised basic measurement $\vect{M}_B$ from $\vect{M}$ to leave a valid subnormalised $\vect{M}' = \vect{M}-\vect{M}_B$. In this case at least one of the following must also be true:

\begin{enumerate}

\item There exists an $x^*$, and a set  $y_a$, such that $M(ab|x^* y_a)>0 \; \forall \, a,b$. It is possible to subtract the subnormalised basic measurement $M_B(ab|xy)=k \delta_{x, x^*}\delta_{y, y_a}$ from $\vect{M}$, where $k=\min_{a,b} M(ab|x^* y_a)$, generating at least one additional 0 element in $\vect{M}'$. \label{wiring1}

\item There exists a $y^*$, and a set $x_b$, such that $M(ab|x_b y^*)>0 \; \forall \, a,b$. It is possible to subtract the subnormalised basic measurement $M_B(ab|xy)=k \delta_{y, y^*}\delta_{x, x_b}$ from $\vect{M}$, where $k=\min_{a,b} M(ab|x_b y^*)$, generating at least one additional 0 element in $\vect{M}'$. \label{wiring2}

\end{enumerate}

\item It is not possible to subtract a (non-zero) subnormalised basic measurement from $\vect{M}$ without creating a negative component. In this case there exist sets $a_x$, and $b_{xy}$, such that $M(a_x b_{xy}  |xy)=0 \; \forall \, x,y$, and there exist sets $b_y$, and $a_{xy}$, such that $M(a_{xy} b_y  |xy)=0 \; \forall \, x$ \label{nowiring2}. In terms of the representation of $\vect{M}$ as an array, this means that each row of blocks (corresponding to fixed $x$) contains a row of components (corresponding to fixed $x$ and $a=a_x$) with at least one zero entry in each block (where $b=b_{xy}$). Similarly, each column of blocks contains a column of components with at least one zero entry in each block.

\end{enumerate}

But case~\ref{nowiring2} is impossible. The example at the end of
this subsection may be helpful in understanding the various stages
of the following.

To derive a contradiction, suppose that case~\ref{nowiring2} holds.
Consider a value $x^*$, and the corresponding $a_{x^*}$ and
$b_{x^*y}$ for which $M(a_{x^*} b_{x^*y}  |x^* y)=0 \; \forall y$.
Consider a product state $\vect{P}$ such that $P(a_{x^*} b_{x^*y}  |
x^*  y)=1 \; \forall y$, and another product state $\vect{P}'$
obtained from $\vect{P}$ by swapping the outputs $a=a_{x^*}$ with
$a=a'$ when  $x=x^*$ (a local relabelling of the outputs).

From Equation~(\ref{dynamic_eqn}) and the definition of a
subnormalised total measurement array, it is clear that
\begin{equation}
\vect{M}.(\vect{P} - \vect{P}') = \sum_{a,b,x,y} M(ab|xy) (P(ab|xy)
- P'(ab|xy) ) = 0. \label{diff_eqn}
\end{equation}
This implies that
\begin{equation}
\sum_{y} (M(a_{x^*} b_{x^* y}  |x^* y)- M(a' b_{x^* y}  |x^* y)) =
0,
\end{equation}
hence $M(a' b_{x^* y}  |x^* y)=0 \; \forall y$. By modifying the
chosen parameters $x^*$ and $a'$, this generalises to the result
that  $M(a b_{xy}  |x y)=0 \;  \forall \, a, x, y$. Similarly,
$M(a_{xy} b  |x y)=0 \; \forall \, b,x, y$. In terms of the array,
each subarray, corresponding to particular values of $x$ and $y$,
contains both a row and a column of zeroes.

But now consider an arbitrary element $M(a^* b^*| x^* y^*)$. Let
$\vect{P}$ be a bipartite product state with $P(a^* b^*| x^* y^*)=1$
and $P(a^* b_{x^* y} |x^* y)=1 \; \forall \, y \neq y^*$, and
$\vect{P}'$ a bipartite product state obtained from $\vect{P}$ by
swapping  $a=a^*$ with  $a=a_{x^* y^*}$ when  $x=x^*$. From
Equation~(\ref{diff_eqn}), $M(a^* b^*| x^* y^*) = 0$. This
contradicts the initial assumption that $\vect{M}$ is nonzero. Hence
case~\ref{nowiring2} is impossible, and either case~\ref{wiring1} or
case~\ref{wiring2} must hold. Iteratively subtracting subnormalised
basic measurements from $\vect{M}$ until the zero vector remains
yields an expression for $\vect{M}$ as a convex combination of basic
measurements.

We now illustrate the above proof for an example bi-partite
measurement, where each subsystem is characterised by two fiducial
measurements each with three outputs. Consider a situation in which
case \ref{nowiring2} holds (it is not possible to remove a basic
measurement), and the representation of $\vect{M}$ as an array
contains zeroes as follows (with all other elements unknown):
\begin{equation}
\left( \begin{array}{c|c}  \begin{array}{ccc} 0&.&. \\ |&.&. \\
|&0&- \end{array}  &  \begin{array}{ccc} .&|&. \\ .&0&.
\\ \!-\!&-&0 \end{array} \\ \hline  \begin{array}{ccc} |&.&.
\\ 0&-&- \\ |&.&. \end{array}  &  \begin{array}{ccc}
.&|&. \\ 0&|&- \\ .&0&. \end{array}  \end{array} \right)
\end{equation}

Here, the outer matrix corresponds to the measurement choice ($x$
vertically, $y$ horizontally), and the inner submatrices correspond
to the outcomes ($a$ vertically, $b$ horizontally). By comparing
pairs of product states as above, deduce that each line of zeroes
can be extended perpendicularly, yielding a row and column of zeroes
in each submatrix:
\begin{equation}
\left( \begin{array}{c|c}  \begin{array}{ccc} 0&0&0 \\ .&0&. \\
.&0&. \end{array} &  \begin{array}{ccc} .&.&0 \\ 0&0&0
\\ .&.&0 \end{array} \\ \hline \begin{array}{ccc} 0&.&. \\
0&0&0 \\ 0&.&. \end{array}  &  \begin{array}{ccc} 0&.&.
\\ 0&.&. \\ 0&0&0 \end{array}  \end{array} \right)
\end{equation}

It then follows that all of the unknown elements must also be zero.
For example, in order to show that $M(1 0| 0 0)=0$ (the element
below the top left element), consider the states $\vect{P}$ and
$\vect{P}'$ as follows (non-zero elements of $\vect{P}$ are shown
boxed and non-zero elements of $\vect{P}'$ are shown in bold):
\begin{equation}
\left( \begin{array}{c|c}  \begin{array}{ccc} \textbf{1}&0&0 \\
\fbox{1}&0&0 \\ 0&0&0 \end{array}  &
\begin{array}{ccc} 0&0&\textbf{1} \\ 0&0&\fbox{1}  \\ 0&0&0
\end{array} \\ \hline  \begin{array}{ccc}
\fbox{\textbf{1}}&0&0 \\ 0&0&0 \\ 0&0&0 \end{array}  &
\begin{array}{ccc} 0&0&\fbox{\textbf{1}} \\ 0&0&0 \\ 0&0&0
\end{array}  \end{array} \right)
\end{equation}

\subsection{Proof of Theorem~\ref{3systemtheorem}}\label{counterexample}

We prove Theorem~\ref{3systemtheorem} with an explicit
counterexample - that is, a joint measurement on three subsystems,
which cannot be considered as a convex combination of basic
measurements. The subsystems are characterised by two binary-outcome
fiducial measurements each. The joint measurement has 8 possible
outcomes, $\vect{R}_0,\ldots,\vect{R}_7$, with
\begin{eqnarray}
R_0(001|000)=R_1(110|000)=1 \label{3box1_eqn}\\
R_2(000|100)=R_3(100|100)=1\\
R_4(101|010)=R_5(111|010)=1\\
R_6(010|001)=R_7(011|001)=1, \label{3box4_eqn}
\end{eqnarray}
and all other components $0$. This measurement is rather subtle.
Note that each individual effect here is in fact a product effect,
in keeping with Corollary~\ref{noenteffect}, which states that there
are no entangled effects. Nevertheless the measurement as a whole
cannot be realised as a basic measurement or convex combination
thereof.

\begin{figure}
\includegraphics[width=5cm]{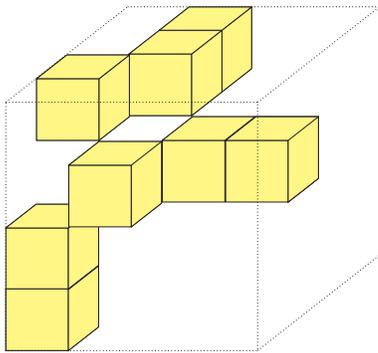}
\caption{This diagram shows the three-dimensional array
corresponding to the total measurement array of a non-basic
tri-partite measurement. The shaded and unshaded entries have values
1 and 0 respectively. Note that the shaded 2x1x1 rectangles can each
be slid parallel to their long edge to complete the 2x2x2 cube in
the front upper left of the array (corresponding to performing
fiducial measurements x=0 on each subsystem). The invariance of the
total measurement array under such transformations is a consequence
of the no-signalling conditions.}\label{3boxmeasurement}
\end{figure}

To see that this cannot be constructed from basic measurements,
consider the total measurement array $\vect{M}$, illustrated in
Figure~\ref{3boxmeasurement}. Note that element  $M(001|000)$ is
surrounded by 3 zeroes in the block with $x=y=z=0$ (i.e. $M(101|000)
= M(011|000) = M(000|000)=0$). As the total measurement array for a
basic measurement is insensitive to the output of the last subsystem
measured, it consists of lines of 1s inside blocks. A basic
measurement cannot therefore have non-zero $M(001|000)$ without also
having nonzero $M(101|000), M(011|000)$ or $M(000|000)$. Given that
all the convex coefficients and basic measurements must be positive,
it is therefore impossible to construct $M(abc|xyz)$  from convex
combinations of basic measurements. Since each non-zero element of
$M(abc|xyz)$ corresponds to a different measurement outcome, there
are no equivalent vectors which implement the same measurement,
hence this measurement is impossible to simulate by any convex
combination of basic measurements.

It remains only to show that $\{ \vect{R}_r \}$ does indeed
represent a valid measurement. The condition that
$\vect{R}_r.\vect{P}\geq 0$ for all valid $\vect{P}$ is ensured by
the positivity of the entries of $\vect{R}_r$. In addition, outcome
probabilities should sum to $1$, meaning that
\begin{equation}
\sum_r \vect{R}_r.\vect{P} = \vect{M} . \vect{P} =
\sum_{a,b,c,x,y,z} M(abc|xyz) P(abc|xyz) = 1 \label{D_eqn},
\end{equation}
for all states $\vect{P}$ satisfying the positivity, normalisation
and the no-signalling conditions. To see that this is indeed the
case, evaluate the sum in Equation~(\ref{D_eqn}) to obtain
\begin{equation}  \label{ok_measurement}
\vect{M} \cdot \vect{P} =
\begin{array}{l}
 P(001|000) + P(110|000)  \\
+ P(000|100) + P(100|100) \\
+ P(101|010)+ P(111|010)\\
+ P(010|001) + P(011|001)
\end{array}
\end{equation}
The no-signalling conditions ensure that
\begin{eqnarray}
P(000|100) + P(100|100) &=& P(000|000) + P(100|000) \nonumber \\
P(101|010)+ P(111|010) &=& P(101|000)+ P(111|000) \nonumber \\
P(010|001) + P(011|001) &=& P(010|000) + P(011|000) \nonumber
\end{eqnarray}
Substituting in Equation~(\ref{ok_measurement}), and using the
normalisation of $\vect{P}$,
\begin{equation}
\vect{M} \cdot \vect{P} = \sum_{a,b,c} P(abc|000) = 1.
\end{equation}

Hence $\{ \vect{R}_r \}$ represents a valid tri-partite measurement
that cannot be simulated by a convex combination of basic
measurements.

\subsection{Proof of Theorem~\ref{postselecttheorem}}

Although joint measurements on three or more subsystems in box world
cannot generally be implemented using fiducial measurements,
Theorem~\ref{postselecttheorem} implies that they are still in some
sense simple, as they can be simulated using local fiducial
measurements and post-selection.

To simulate a general measurement described by $\{\vect{R}_r\}$,
first perform a random fiducial measurement $\mbf{x}$ on the
complete system (composed of a random fiducial measurement $x_1,
x_2, \ldots , x_N$ on each subsystem), in which each $\mbf{x}$
occurs with constant probability $q$. If result $\mbf{a}$ is
obtained in the fiducial measurement, then the general measurement
outcome $r$ is given with probability
\begin{equation}
\mathrm{Pr}(r|\mbf{a}\mbf{x}) =
\frac{R_r(\mbf{a}|\mbf{x})}{\max_{\mbf{ax}} M( \mbf{a}|\mbf{x}) }
\end{equation}
and `failure' is declared with probability
\begin{eqnarray}
\mathrm{Pr}(\mathrm{fail}|\mbf{a}\mbf{x}) &=& 1-\sum_r \mathrm{Pr}(r|\mbf{a}\mbf{x})) \nonumber \\
    &=& 1-  \frac{M(\mbf{a}|\mbf{x})}{\max_{\mbf{ax}} M( \mbf{a}|\mbf{x}) }
\end{eqnarray}
Note that due to the way they are constructed, and the fact that
$R_r(\mbf{a}|\mbf{x})\geq 0$, these probabilities are all positive,
and the total probability for declaring failure or some output is
one.

The probability of obtaining output $r$ given that the simulated
measurement succeeds is then given by
\begin{eqnarray}
\mathrm{Pr}(r|\mathrm{success}) &=& \frac{\sum_{\mbf{a x}} q P(\mbf{a}|\mbf{x}) \mathrm{Pr}(r|\mbf{a}\mbf{x}) }{\sum_{r'} \sum_{\mbf{a x}} q P(\mbf{a}|\mbf{x}) \mathrm{Pr}(r'|\mbf{a}\mbf{x}) }  \nonumber \\
&=& \sum_{\mbf{a x}}R_r(\mbf{a}|\mbf{x})P(\mbf{a}|\mbf{x})
\end{eqnarray}
which is exactly what one would expect from a perfect implementation
of the measurement. We have therefore proved
Theorem~\ref{postselecttheorem}.

The same approach does not apply to quantum theory because there
$R_r(\mbf{a}|\mbf{x})$ can contain negative components. In fact, the
theorem does not hold for quantum theory. Bell measurements, for
example, cannot be simulated by local Pauli measurements and
post-selection.

\section{Consequences for information processing}\label{consequences}

One reason for investigating the available measurements in box world
is that we can draw conclusions about information processing in box
world, and contrast this with information processing in quantum
theory. The facts that there are no entangled effects in box world,
that measurements are limited to probabilistic mixtures of basic
measurements (for single and bipartite systems), and can be
simulated with postselection (for any system) imply that information
processing in box world is in some ways rather limited. This is
\emph{despite} the fact that box world allows highly entangled
states, which can exhibit strong nonlocality in violation of
Tsirelson's bound.

Consider first entanglement swapping \cite{swapping}. In quantum
theory, the simplest example of entanglement swapping is as follows.
Alice and Bob share two quantum systems in a singlet state $ |
\psi_{-} \rangle_{AB_1} $, and Bob and Charlie share two more
systems, also in a singlet state $ | \psi_{-} \rangle_{B_2C} $. Bob
performs a Bell basis measurement on systems $B_1$ and $B_2$ and
announces the outcome. Alice's and Charlie's systems will now be in
a maximally entangled state (where which entangled state they share depends on Bob's outcome). 
Refs.~\cite{barrett} and \cite{coupler} both offer proofs
that an analogous procedure is impossible in box world in the
special case that each system is characterised by two binary-output
fiducial measurements. Here we show that this result is general.
\begin{theorem}
In box world, there is no entanglement swapping. In particular,
suppose that Alice shares with Bob any number of systems in a joint
state $\vect{P}$, and Bob shares with Charlie any number of systems
in a joint state $\vect{Q}$ and that the initial joint state of all
systems is a direct product of $\vect{P}$ and $\vect{Q}$. Then there
is no measurement that Bob can perform on his systems that will, for
some outcome, result in an entangled state shared between Alice and
Charlie.
\end{theorem}

{\bf Proof.} Let the systems held by the parties be denoted $A$,
$B_1$, $B_2$ and $C$, such that $\vect{P}$ is the state of $A$ and
$B_1$ and $\vect{Q}$ is the state of $B_2$ and $C$. The proof is
general enough that any of these may themselves be composite
systems. Let $\vect{P}_{\mbf{b_1}\mbf{y_1}}$ be the collapsed state
of the $A$ system, conditioned on fiducial measurement $\mbf{y_1}$
being performed on the $B_1$ system with outcome $\mbf{b_1}$. The
collapsed state is defined such that its components satisfy
\begin{equation}
P_{\mbf{b_1}\mbf{y_1}}(\mbf{a}|\mbf{x}) =
\frac{P(\mbf{a}\mbf{b_1}|\mbf{x}\mbf{y_1})}{\sum_{\mbf{a}}
P(\mbf{a}\mbf{b_1}|\mbf{x}\mbf{y_1})}.
\end{equation}
If $\vect{Q}$ is a joint state of systems $B_2$ and $C$, then
$Q_{\mbf{b_2}\mbf{y_2}}$ is a collapsed state of the $C$ system,
defined similarly.

Suppose that Bob makes a joint measurement on all of his subsystems,
with an outcome $r$, and that Alice and Charlie perform fiducial
measurements $\mbf{x}$ and $\mbf{z}$ with outcomes $\mbf{a}$ and
$\mbf{c}$. It is quite easy to show that the probability of getting
outcomes $r, \mbf{a},\mbf{c}$ is given by
\begin{equation}
\mathrm{Pr}(r \mbf{a c} | \mbf{x z})\! = \!\sum_{ \mbf{b_1 y_1}
\atop \mbf{b_2 y_2}} R_r(\mbf{b_1 b_2}|\mbf{y_1 y_2}\!)
 P(\mbf{a}\mbf{b_1}|\mbf{x}\mbf{y_1}\!)Q(\mbf{b_2}\mbf{c}|\mbf{y_2}\mbf{z}\!).\label{jointprob}
\end{equation}
Equation~(\ref{jointprob}) can be reexpressed as
\begin{equation}
\mathrm{Pr}(r \mbf{a c} | \mbf{x z})=\sum_{\mbf{b_1 b_2 y_1 y_2}}
C^r_{\mbf{b_1 b_2 y_1 y_2}} P_{\mbf{b_1 y_1}}(\mbf{a}|\mbf{x})
Q_{\mbf{b_2 y_2}}(\mbf{c}|\mbf{z}),
\end{equation}
with
\begin{equation}
C^r_{\mbf{b_1 b_2 y_1 y_2}} = R_r(\mbf{b_1 b_2}|\mbf{y_1 y_2})
P(\mbf{b_1}|\mbf{y_1}) Q(\mbf{b_2}|\mbf{y_2}).
\end{equation}

The collapsed state of the $AC$ system, given outcome $r$ for Bob's
measurement, satisfies
\begin{equation}
P_{r}(\mbf{a c} | \mbf{x z}) = \frac{\mathrm{Pr}(r \mbf{a c} |
\mbf{x z})}{ \mathrm{Pr}(r|\mbf{x z})} = \frac{\mathrm{Pr}(r \mbf{a
c} | \mbf{x z})}{\sum_{\mbf{a c}} \mathrm{Pr}(r \mbf{a c}|\mbf{x
z})}
\end{equation}
from which it follows that
\begin{equation}
P_{r}(\mbf{a c} | \mbf{x z})=\sum_{\mbf{b_1 b_2 y_1 y_2}}
\lambda^r_{\mbf{b_1 b_2 y_1 y_2}} P_{\mbf{b_1 y_1}}(\mbf{a}|\mbf{x})
Q_{\mbf{b_2 y_2}}(\mbf{c}|\mbf{z})  \label{lhv_eqn}
\end{equation}
where
\begin{equation}
\lambda^r_{\mbf{b_1 b_2 y_1 y_2}} = \frac{C^r_{\mbf{b_1 b_2 y_1
y_2}}}{\sum_{\mbf{b_1 b_2 y_1 y_2}} C^r_{\mbf{b_1 b_2 y_1 y_2}}}.
\end{equation}
Due to the positivity of $R_r(\mbf{b_1 b_2}|\mbf{y_1 y_2})$
(Theorem~\ref{positivity}), the $\lambda^r_{\mbf{b_1 b_2 y_1 y_2}}$
are all positive. Note also that $\sum_{\mbf{b_1 b_2 y_1 y_2}}
\lambda^r_{\mbf{b_1 b_2 y_1 y_2}}=1$. Thus Equation~(\ref{lhv_eqn})
represents a separable state for Alice and Charlie. Hence Bob's
measurement cannot introduce entanglement between Alice and Charlie,
whatever the result. $\square$

\begin{corollary}
In box world, states cannot be teleported.
\end{corollary}
This is immediate given the impossibility of swapping entanglement.
If teleportation were possible in box world, then it would be
possible to achieve entanglement swapping by teleporting one half of
an entangled state.

Note that in \cite{skipchick1, skipchick2}, it is shown that
entanglement swapping is possible in an alternate theory within the
probabilistic framework, that has a smaller state space than box
world \footnote{For example, the bipartite state space in
\cite{skipchick1} has $a_1, a_2, x_1, x_2 \in \{0,1\}$ and is the
convex hull of all local probability distributions and the PR-box
distribution given by (\ref{PS-state_eqn}). It excludes other
entangled non-signalling distributions}. This is a good illustration
of the trade off between states and measurements in probabilistic
theories.

Our final theorem concerns dense coding. In quantum theory, dense
coding allows Alice to send two bits of classical information to Bob
via the transmission of only one qubit, provided that they intially
share an entangled state \cite{dense}. This contrasts the fact that
if no entanglement is shared, a single qubit can only be used to
transmit one bit of classical information \cite{holevo}. The
procedure is as follows. Suppose that Alice and Bob share two qubits
in a singlet state $|\psi_{-}\rangle_{AB}$. Alice now performs one
of four possible unitary transformations, $I$, $\sigma_x$,
$\sigma_y$, $\sigma_z$ on her qubit, depending on the two classical
bits she wishes to send. She sends the qubit to Bob, who, with the
two qubits in his possession, performs a Bell basis measurement. The
outcome of this measurement tells him with certainty which
transformation Alice performed.
\begin{theorem}
In box world, there is no dense coding.
\end{theorem}
{\bf Proof.} Suppose that Alice and Bob initially share a bipartite
system in a joint state $\vect{P}$. In a dense coding protocol,
Alice would perform a transformation $\vect{T}$ on her system, where
$\vect{T}$ depends on the message she wishes to send. Recalling
Equation~\ref{dynamicsequation} for single systems, the effect of
Alice's transformation on the global state is
$\vect{P}\rightarrow\vect{P}'$ where
\begin{equation}
P'(a'b|x'y) = \sum_{ax} T(a'|x',a|x)P(ab|xy).
\end{equation}
Alice sends her system to Bob, who performs a measurement on the
bipartite system. However, by Theorem~\ref{wiringtheorem}, Bob's

measurement is a convex combination of basic measurements. If Bob is
to learn the message with certainty, random choices cannot help, so
assume his measurement is a basic measurement. There are two cases:
either the basic measurement begins with a fiducial measurement on
system $A$ or it begins with a fiducial measurement on system $B$.
In the second case, Bob could equally have performed the fiducial
measurement on system $B$ before Alice sends system $A$, indeed
before she performs her transformation. Hence the protocol is
equivalent to a protocol with no entanglement, in which Alice simply
begins with a single system, performs a transformation and sends it
to Bob. The amount of classical information transmitted can be no
more than the sending of a single box allows. In the first case, the
proof is slightly more involved. Consider the fiducial measurement
on system $A$ that Bob is to perform. In an equivalent protocol,
Alice performs this measurement herself, just after her
transformation, and then sends to Bob the classical outcome of the
measurement, instead of system $A$. Let this measurement have $d$
possible outcomes. In any protocol in which the only transmission
from Alice to Bob is a number from $1$ to $d$, it is impossible for
Bob to distinguish $>d$ messages, even if there is pre-shared
entanglement.\footnote{Suppose that there were such a protocol. Then
there is another protocol involving no transmission, in which Bob
simply guesses the number $1$ to $d$ that would have been sent, and
infers a guess for the message. In this second protocol, Bob guesses
one of $>d$ messages correctly with probability at least $1/d$, in
violation of the no-signalling principle.} But even without
pre-shared entanglement, the transmission of system $A$ would
suffice for Bob to distinguish $d$ possible messages. Alice simply
encodes the message into the outcome of the $d$-outcome fiducial
measurement. Hence the dense coding protocol confers no advantage.
$\square$

\section{Conclusions}

Box world is not classical, nor quantum, but it has a natural and
easy definition in terms of an operational framework. Of course it
does not describe our universe. So why bother with it? Simply for
the sake of comparison with quantum theory. In particular, it is
interesting to explore the information processing possibilities of
box world and to compare these with the possibilities in quantum
theory. There are ways in which box worlders are better off than the
inhabitants of a quantum universe: as van Dam showed \cite{van-dam},
they find communication complexity problems trivial. But in other
ways, the box worlders are worse off: the results of this paper
imply that they cannot do entanglement swapping, teleportation or
dense coding.

These differing powers can be traced back to a tradeoff that exists
between the states which a theory allows and the measurements it
allows. Box world is permissive with respect to states - all
nonlocal correlations can be realized. But this forces a paucity of
measurements which means that the theory is very restricted in other
ways. Quantum theory is actually remarkable in the balance it
achieves between the two, yielding potent nonlocal correlations in
addition to a broad range of possible measurements and dynamics.
This leads us to speculate that quantum theory is, in at least some
ways, optimal.

Finally, one issue that we have not raised is that of computation.
It is possible to define a circuit model for computation in box
world, similar to the classical and quantum circuit models
\cite{barrett}. Would a box computer be even more powerful than a
quantum computer? Alternatively, in view of the restricted dynamics,
would it perhaps be no better than a classical computer?

\textbf{Acknowledgments} The authors thank Andreas Winter for a
helpful discussion on dense coding. AJS acknowledges support from a
Royal Society University Research Fellowship, and also support from
the U.K.~EPSRC ``QIP IRC' project' whilst working on this paper at
the University of Bristol. Part of this work was done whilst JB was
supported by an HP Fellowship. JB is currently supported by an EPSRC
Career Acceleration Fellowship. This work was supported in part by
the EU's FP6-FET Integrated Projects SCALA (CT-015714) and QAP
(CT-015848).

\end{document}